\documentclass[amsmath,eqsecnum,preprint,pre,aps,showpacs]{revtex4}
\usepackage{amsmath}
\usepackage{graphicx}
\usepackage{dcolumn}
\usepackage{bm}
\usepackage{float}
\usepackage{xcolor}
\usepackage{amssymb}

\begin{document}


\title{Superheating and melting phenomena of a vibrated granular layer of cubic particles.}
\author{Francisco L\'opez-Gonz\'alez$^{1}$, Gustavo M. Rodr\'iguez-Li\~n\'an$^{2}$, Fernando Donado$^{1}$, Felipe Pacheco V\'azquez$^{3}$ and Luis Fernando Elizondo-Aguilera$^{3,*}$}

\address{$^1$Instituto de Ciencias B\'asicas e Ingenier\'{\i}a de la Universidad Aut\'onoma del Estado de Hidalgo-AAMF,\\Pachuca 42184, Hidalgo, Mexico}
\address{$^2$ Investigadores por M\'exico -- Instituto de Geociencias ,  Universidad Nacional Aut\'onoma de M\'exico, Blv. Juriquilla 3001, 76230 Quer\'etaro, M\'exico}
\address{$^3$ Instituto de F\'isica, Benem\'erita Universidad 
Aut\'onoma de Puebla, Apartado Postal J-48, 72570, Puebla, M\'exico.}
\date{\today}

\begin{abstract}
We report the combined results of both, experiments and molecular dynamics simulations, carried out to investigate superheating phenomena in vertically vibrated granular matter. Specifically, we consider a system of cubic particles, densely packed in a squared-lattice array and subjected to different shaking strengths $\Gamma$, in order to approach a critical value $\Gamma_c$. Below $\Gamma_c$, the superheated crystalline array remains stable indefinitely, whereas above $\Gamma_c$, it transitions progressively into a granular liquid-like state, during a $\Gamma$-dependent timescale $\tau$. We show that, while an increase in frictional contacts substantially lengthens the lifetime of the superheated crystalline state, it does not play a major role in determining the scaling laws describing the dependence of the superheated crystal lifetime on the shaking strength $\Gamma$. Our findings also show that the transition from the superheated-solid to the liquid state of the vibrated system is well captured by a Kolmogorov-Johnson-Mehl-Avrami equation, routinely employed to describe phase transformations in thermal systems.

\end{abstract}


\maketitle

\section{Introduction}

Under certain conditions, some materials remain solid even when heated above the melting temperature $T_m$. This behavior, commonly referred to as superheating in condensed matter physics \cite{lu,mei,wunderlich}, is observed in a variety of materials with very different microscopic characteristics, ranging from ice \cite{iglev, celik,fanetti}, metallic alloys \cite{martynyuk,forsblom,sun,siwick}, polymers \cite{toda1,toda2} and seminconductors \cite{casey} among others \cite{wunderlich}, thus suggesting its universality. Instead of the first-order transition expected above the melting point, a variety of experimental data show the existence of a small temperature domain at which metastable solid states can prevail over a transient time-window. Above such domain, however, the solid phase simply looses its stability and melts all of a sudden \cite{lu,martynyuk,siwick}. Despite a long standing debate \cite{cotterill,forsblom}, a conclusive physical explanation of this phenomena remains elusive.

Superheated solid states, however, are notoriously difficult to achieve experimentally, since the liquid–solid (or gas–solid) interface typically triggers a precipitous melting (or sublimation), thus preventing the crystal from sustaining even slight superheating temperatures. A well-established protocol to circumvent these practical constrains consists in coating the solid substrate with a different material of larger melting point, providing more stability to the crystalline phase of the former, and increasing hence the lifetime of the corresponding superheated states. Concrete realizations of this protocol have been reported, for instance, using antifreeze proteins in solid ice \cite{celik}, coating Ag crystals with a thick Au layer \cite{daeges} or using GaAs coatings on Ge crystals \cite{casey}.   

Although much more scarcely, superheating phenomena has also been observed in driven granular matter, which often mimics thermal behaviors observed at the atomic, molecular and mesoscopic scales in spite of the pervasiveness of energy dissipation  \cite{elizondo,jaime,garzo,zhang,vega}. A hexagonal closed packed monolayer of spherical grains, for example, subjected to sinusoidal vibrations $z(t)=A\sin(2\pi ft)$ -- with $z$ being the vertical position of a container at time $t$ -- is able to maintain its crystalline configuration for an indefinite amount of time for peak accelerations slightly larger than gravity, or $g$ \cite{pacheco1}. For such a system, in which the shaking strength $\Gamma\equiv A(2\pi f)^2/g$ is a control parameter that plays the role of an effective granular temperature \cite{reis,perera}, one finds a critical value $\Gamma^{sph}_c\approx1.38$ separating two different (an mutually exclusive) physical scenarios, namely: (i) for $\Gamma<\Gamma^{sph}_c$ the excited hexagonal array of spheres remains stable indefinitely, with each particle moving vigorously in its lattice position while describing erratic (i.e. non-coordinated) jumps with respect to its neighbors; and (ii) for $\Gamma>\Gamma^{sph}_c$ the system sublimates catastrophically into a gas after a $\Gamma$-dependent timescale $\tau^{sph}$, with the later becoming increasingly smaller for larger $\Gamma$. Thus, metastable states that resemble qualitatively those observed in typical superheated crystalline materials, can also be obtained at the granular level. 

Surprisingly, however, the study of superheating in driven granular matter has remained virtually neglected. To the best of our knowledge, excepting for the investigation reported in Ref. \cite{pacheco1} no additional studies have been carried out so far to elucidate further this phenomenology and, consequently, many relevant aspects remain unresolved. One of them, in particular, refers to the possibility to achieve longer-lasting superheated states, in qualitative similitude to the aforementioned coating protocols applied to conventional crystalline solids. Extending the durability of these transient states would allow to investigate in more detail the metastability of an excited granular crystal and, more crucially, to rationalize further relevant thermodynamic notions, such as phase-transformations and coexistence in a highly dissipative many-body system. This, in turn, may allow for additional analogies between the behavior of driven granular matter and thermal systems to be drawn. 

For these reasons, here we report the results of both, experiments and molecular dynamics (MD) simulations carried out to investigate superheating behavior in a quasi-two-dimensional granular system, comprised by cubic particles that are subjected to periodic mechanical stimuli. These particles where initially organized in a squared-lattice monolayer inside a commensurable container, and subjected to vertical oscillations at different values of $\Gamma$ afterwards. For such a system, in which the energy dissipation mechanisms due to friction are clearly more efficient with respect to a crystalline array of spherical grains, we analyzed different superheated states, and found important qualitative and quantitative differences with respect to the features previously highlighted for a hexagonal monolayer of spherical particles, with a single point contact. 

First, we observed a notorious expansion of the $\Gamma$-domain where superheated states can be obtained, thus implying a shift of the critical shaking strength towards a significantly larger value, $\Gamma_c\approx3.225$. Secondly, approaching $\Gamma_c$ from above, we also found a remarkable increase of the lifetime $\tau_c$ for the superheated states. Furthermore, in our present system we do not find any evidence of a sudden sublimation of the solid phase into a granular gas, as it is observed in the spherical case \cite{pacheco1}. Instead, one observes a gradual melting of the crystal into a liquid-like state, as revealed by the kinetics of the radial distribution function of the system, and implying the possibility to obtain solid-liquid coexistence over the time scale $\tau_c$. These characteristics highlight thus the important role played by energy dissipation in the endurance of granular superheating. 

In addition, we empirically determine the dependence of the melting time $\tau_c$ on the shaking strength $\Gamma$, which, strikingly, follows a power law that resembles closely that found in a hexagonal monolayer of spheres \cite{pacheco1}. Finally, we show that the kinetics of melting obtained in both, experiments and simulations, is well captured by the so-called Kolmogorov-Johnson-Mehl-Avrami model, routinely employed to describe the progress of phase transformations in a variety of systems and conditions \cite{kiana}. Therefore, the example considered here provides an amenable model to analyze \emph{phase}-transformations in granular matter. To the best of our knowledge, this analogy, along with all the characteristics described in what follows, has never been highlighted before.   

Our work is organized as follows: In Section \ref{mat_met} we provide the main details of both, the experimental setup, and the MD simulations employed to investigate superheated states in a granular system comprised by cubic particles. In Section  \ref{results} we present and discuss our results. For clarity and methodological convenience, this Section is organized in four Subsections. In Subsection \ref{gen_melt}, we describe first the general characteristics of the process of transformation of a superheated monolayer of cubes into a liquid-like state. In subsection \ref{rdf}, we discuss the same process in terms of the kinetics of the two dimensional radial distribution function $g(r)$, allowing to introduce a \emph{melting} timescale, $\tau$. In Subsection \ref{taut}, we analyze the dependence of $\tau$ on the shaking strength $\Gamma$, allowing for the  empirical determination of the critical value $\Gamma_c$. More crucially, however, we also discuss the connection with previous work, as well as the scaling laws describing superheating in driven granular systems. Furthermore, we elaborate on the geometrical constrains that trigger the transition from a superheated granular solid to the liquid-like state. Subsection \ref{prof} discusses briefly the connection of our findings with the so-called Johnson-Mehl-Avrami-Kolmogorov theory, broadly used for the description of the kinetics of non-equilibrium transformations. Finally, in Sec. \ref{conclu} we provide our concluding remarks.

\section{Materials and Methods}\label{mat_met}

\subsection{Experiments}\label{exp_info}

Our experimental setup is based on previous work \cite{elizondo}. It consisted of $900$ cubic particles (plastic dices) of side length $d = 5\pm 0.1$ mm, mass $m\approx0.123$ g, and friction coefficient (\emph{dice-face} to \emph{dice-face}) $\mu_{DD} = 0.47\pm0.02 $. These particles were poured into an acrylic squared box (friction coefficient \emph{box} to \emph{dice-face}, $\mu_{BD} = 0.35$) of length $L\approx 150.5\pm 0.01$ mm and no top lid. This confinement geometry allows to easily assemble a closed-packed squared-lattice monolayer of cubes at the bottom of the container. The filling fraction $\varphi$ of the system, defined here as the ratio between the projected squared area of the particles at rest, and the total area of the bottom of the box, was $\varphi=0.98\pm.01<1$, which yielded enough space to the particles to move in their lattice sites without being jammed. 

The container was fixed to a shaker (6.5'' KENWOOD speaker) used to vibrate sinusoidally the system as $z(t)=A\sin\left(2\pi ft\right)$, with $z(t)$ being the vertical position of the bottom at time $t$. For practical convenience, the frequency $f$ of the oscillating signal was controlled using two function generators (Pasco Scientific PI-9587C and PI-8127), whose signals were interchanged via a standard 12VDC relay (Songle, SRD-12VDC-SL-C). At each fixed value of $f$, the shaking strength $\Gamma=A(2\pi f)^2/g$ was determined via an accelerometer (Analog Devices, ADXL345 ) attached to the cubic container. The first function generator was employed to produce an initial steady state of the excited crystalline monolayer, vibrated here at $\Gamma_{0}=1.68$ ($f=125$ Hz,) in all the cases. Subsequently, the relay was employed to increase instantaneously the frequency $f$ and, hence, to quench from $\Gamma_0$ towards different values of the shaking strength, denoted in what follows as $\Gamma_f$. 

\begin{figure}[h]
\centering
\includegraphics[width=.99\textwidth, height=0.5\textwidth]{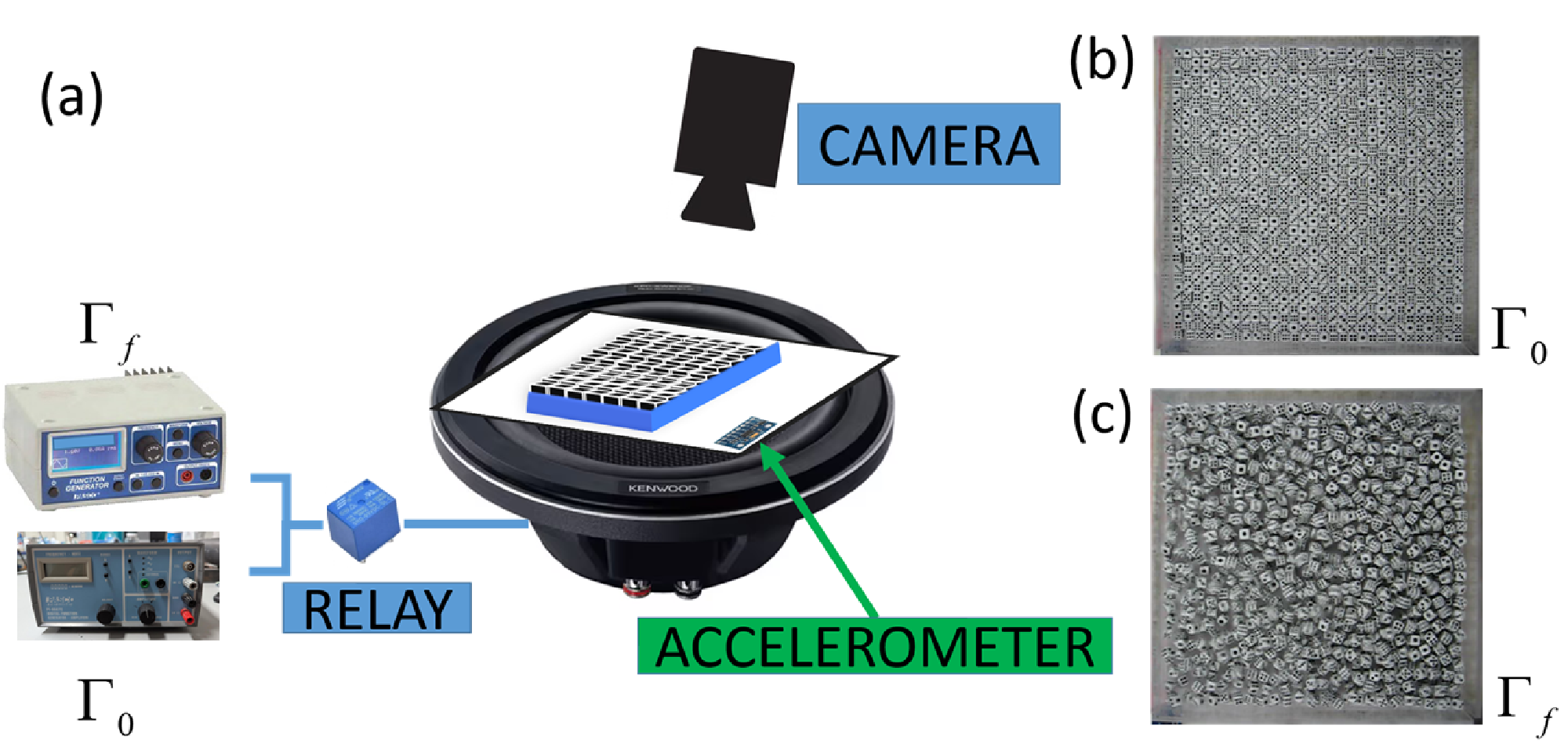}
\caption{(a) Schematic representation of the experimental set up employed to obtain superheated granular monolayers. The experimental samples were vibrated using a shaker and two function generators, interchanged through a relay (see the text). (b) Initial configuration of 900 dices vibrated at $\Gamma_{0}=1.68\,$ in squared-lattice array. (c) Final configuration obtained for 900 dices in a liquid-like state after 108 seconds of vibrations at $\Gamma_{f}=3.36$.} \label{setup}
\end{figure}

The experimental system was filmed from above at 30 frames per second using a standard camera (CANON PowerShot SD750). For each selected value of $\Gamma_f$, 5 realizations of the same experiment were used to report averaged quantities. For reference, a schematic representation of our experimental set up is shown in Fig. \ref{setup}(a). In addition, Figs. \ref{setup}(b) and \ref{setup}(c) illustrate, respectively, the initial excited state of the squared-lattice monolayer of cubic particles ($\Gamma_0=1.68$) and a final liquid state obtained at $\Gamma_f=3.36$ for times of order $10^2$s. 

\subsection{Simulations}\label{simi}

Our simulation protocols also follow from Ref. \cite{elizondo}. The reader is invited to visit Section 2.2 of this reference, which describes in all detail the technicalities involved. As explained there, our simulations are based on a MD custom code of our own creation, in which the cubic particles are modeled as regular hexahedrons of side length $d$, with uniform mass density $\rho_m=m/d^3$ and moment of inertia $I=md^2/6$ (Fig. \ref{simi}a). On the vertices of each hexahedron, eight spheres of radius $r_s=0.1d$ are placed, and bounded together with cylinders of the same radius and length $l_c=0.8d$ (Fig. \ref{simi}b), with the later used to represent the edges of the cubes. Inter-particle interactions were accounted for following the procedure outlined in \cite{elizondo}, which demonstrates that this simplified model system captures efficiently the structural and dynamical behavior of the present experimental system comprised by cubic particles. 

\begin{figure}[h]
\centering
\includegraphics[width=.24\textwidth, height=0.24\textwidth]{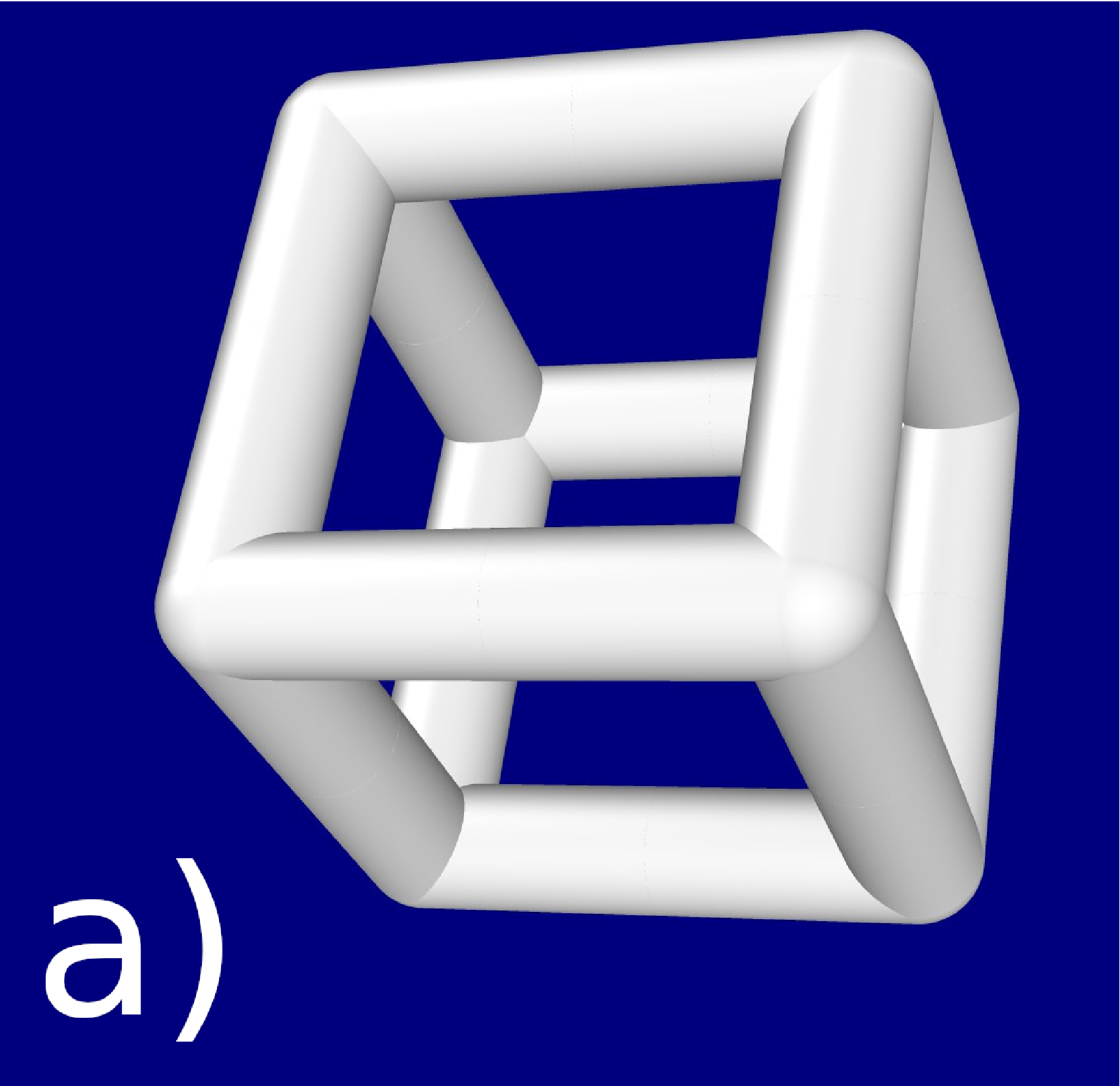}
\includegraphics[width=.24\textwidth, height=0.24\textwidth]{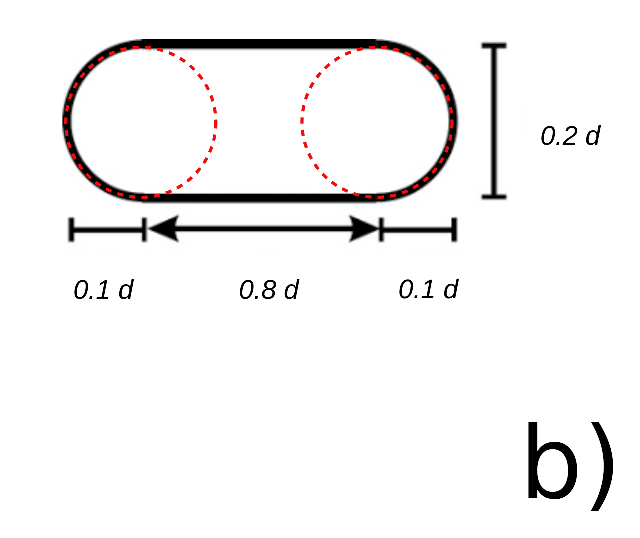}
\includegraphics[width=.24\textwidth, height=0.24\textwidth]{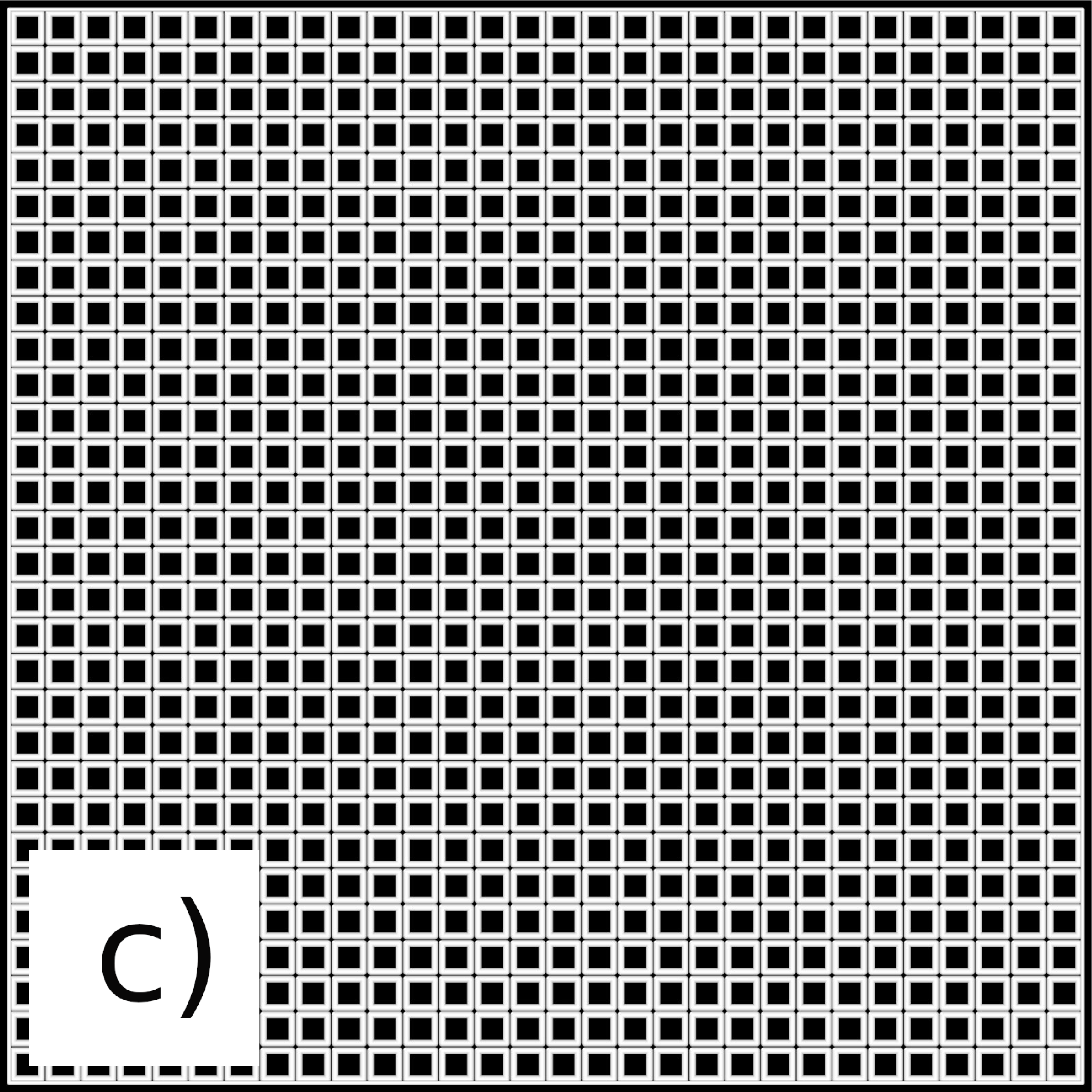}
\includegraphics[width=.24\textwidth, height=0.24\textwidth]{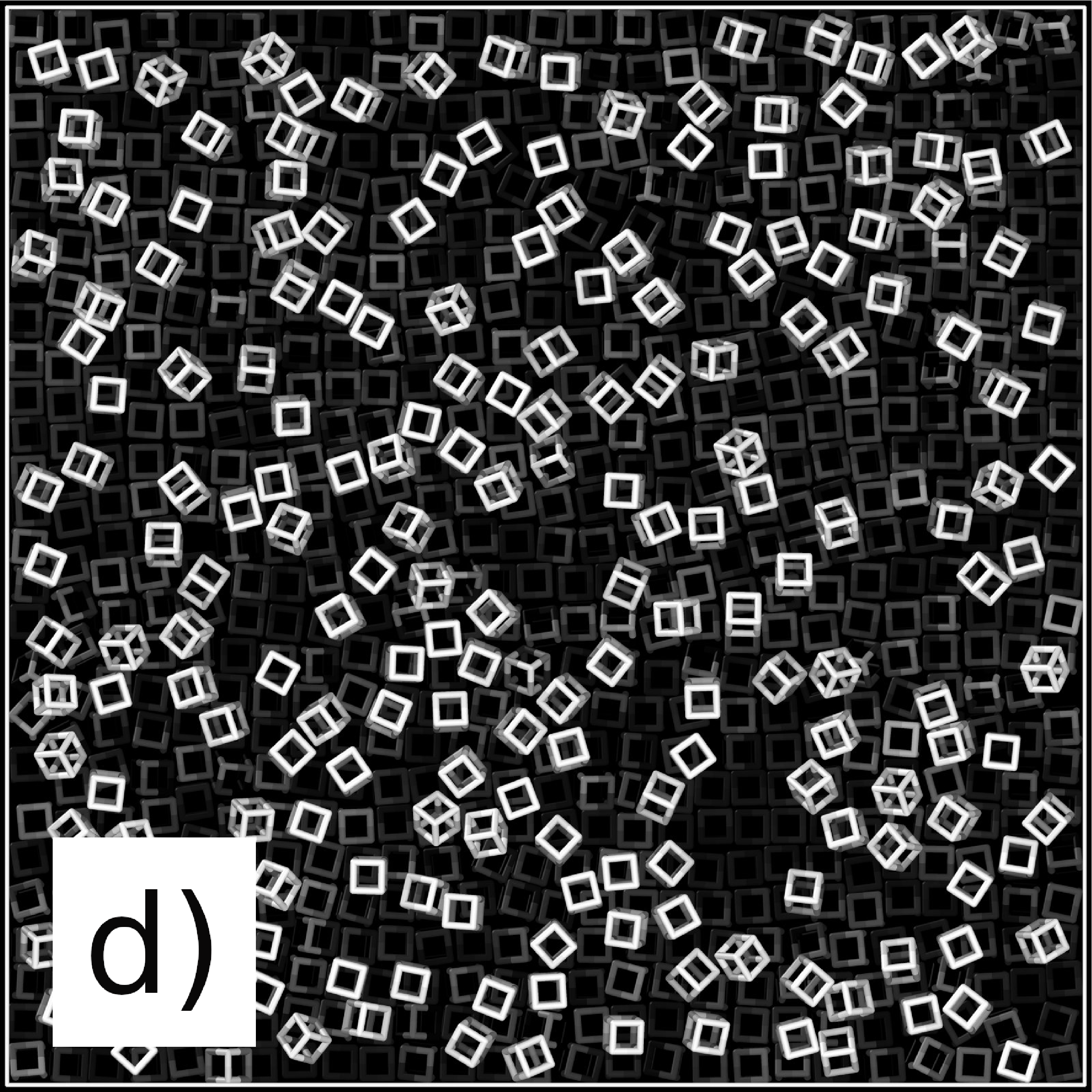}
\caption{(a) Regular hexahedrons used in our MD simulations to represent the cubic particles. (b) Schematic representation of a hexahedron's side, composed by a cylinder of length $l_c=0.8d$ and two spheres of radius $r_s=0.1d$ on the vertex (see also Sec. 2.2 of Ref. \cite{elizondo}). (c) and (d) are snapshots of the initial and final configurations, respectively, obtained in a simulated sample for $\Gamma>\Gamma_c$.} \label{simi}
\end{figure}

In the simulations, the aforementioned particles interact with four vertical planes, located at $x=0$, $x=30d+\lambda$, $y=0$ and $y=30d+\lambda$, which account for the container walls. Here the parameter $\lambda$ was fixed to $\lambda=0.3d$ to prevent jamming. The particles also interact with the horizontal plane, whose vertical position was varied as $z(t)=A\sin(2\pi f t)$. In this case, the oscillation frequency $f$ was fixed to $f=60$Hz, and the amplitude of oscillation $A$ was adjusted accordingly to obtain different values of $\Gamma$. Following Ref. \cite{elizondo}, the interaction between a cube and the container walls was accounted for only by means of the contacts between the spheres on the vertices of a particle and the five planes, using a restitution coefficient $\epsilon_w= 0.8$ and a friction coefficient $\mu_w=0.2$. We should mention here that the numerical values of these parameters were adjusted empirically, in order to obtain time scales and $\Gamma$-values comparable to the experimental samples. 
 
At the beginning of each simulation run, $900$ composed particles were arranged in a squared lattice configuration inside the cubic box of sides $L=30.3d$, with a horizontal center-of-mass separation between particles of $1.001d$, and at a vertical position $z=0.501d$ with respect to the bottom plane of the box. Random initial velocities in the range $[-v_0,v_0]$ were assigned to the particles, following
a constant distribution for all three velocity components, where $v_0=10^{-5}d/t_s$
and with $t_s=10^{-6}$s being the simulation time step. Similarly, we assigned
random initial components for the angular velocity in the range $[-w_0,w_0]$, with
$w_0= v_0/\sqrt{3d}$. All the simulated samples were allowed to evolve over a sufficiently large amount of time, in order to observe the full melting process for large $\Gamma_f$. For each value of the shaking strength, 5 simulation runs were considered to report averaged quantities.

\section{Results}\label{results}

We now discuss the results of the experiments and simulations just described. For this, let us recall first that all the experimental samples studied were initially prepared in a steady state at $\Gamma_0=1.68$, a value of the shaking strength for which we observed that the squared-lattice array of cubic particles remains stable for an indefinite amount of time. This feature, in fact, immediately underscores an important difference with respect to the behavior of a monolayer of spherical grains since, for such a system, this value of the shaking strength already suffices to launch a chain reaction that rapidly yields a granular gas (see, for instance, Figs. 1 and 2 of Ref. \cite{pacheco1}). Thus, a straightforward -- yet relevant -- observation follows: an increase in the contact surface between the constitutive particles provides much more stability to the granular superheated states which. Of course, this is intuitively expected from the more efficient dissipation of energy through larger frictional contributions among the cubic particles (i.e. face-to-face contacts, rather than point-to-point).

As described in Subsection \ref{exp_info}, a relay was used at an arbitrary time ($t=0$) to switch the shaking strength $\Gamma_0$ towards a larger value $\Gamma_f$, and the evolution of the experimental system was monitored afterwards ($t>0$). As explained in the introduction, such a quench leads to one of two possible scenarios for the evolution of the vibrated system: one in which the superheated monolayer of cubes remains stable indefinitely, observed for $\Gamma_f<\Gamma_c$; or another where the ordered array evolves gradually into a granular \emph{liquid}-like state \cite{elizondo}, occurring when $\Gamma_f>\Gamma_c$. We might thus introduce a time scale $\tau_m=\tau_m(\Gamma_f)$, defined as the time elapsed from the instantaneous quench $\Gamma_0\to\Gamma_f$ (at $t=0$), and until the system reaches a stationary \emph{liquid} state. Obviously, for $\Gamma_f<\Gamma_c$, $\tau_m$ is infinite, whereas for $\Gamma_f>\Gamma_c$ it becomes finite. In what follows we shall refer to $\tau_m$ as the \emph{melting} time of the granular system.

\subsection{General characteristics of \emph{melting}}\label{gen_melt}

\begin{figure}[h]
\centering
\includegraphics[width=.9\textwidth, height=0.7\textwidth]{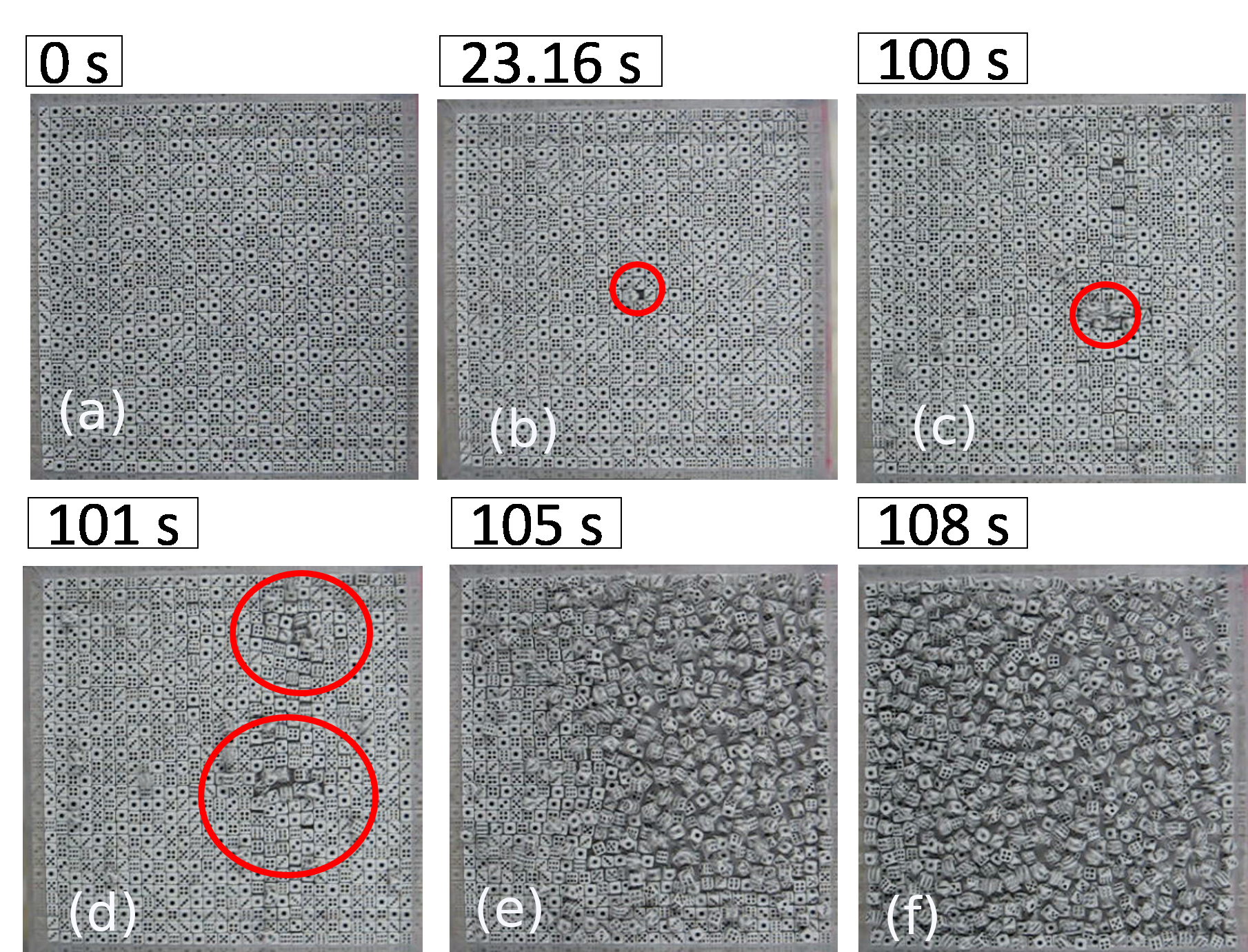}
\caption{Sequence of snapshots illustrating the evolution of the experimental system after an instantaneous quench, from the initial state at shaking strength $\Gamma_0=1.68$ (a), and towards a final value $\Gamma_{f}=3.36$. For times $t<100$s, the squared-lattice array remains stable, with only a few dislocations appearing when individual particles escape spontaneously from the monolayer (red circles in (b) and (c)). At larger times, more particles are able to escape from the monolayer, thus freeing up enough space for an increasing number of particles, which gradually develop random motion and are able to distort their local environment (red circles in (d)). At $t\approx105$s the individual liquid-like clusters already form a larger entity that coexists with the remaining portion of the crystalline structure (e). Finally, at $t\approx108$s the system reaches a liquid-like state (f).} \label{snaps3p36}
\end{figure}

To serve as a reference for the incoming discussion, it is instructive to start by highlighting some generic features of the process of transformation (or \emph{melting}) observed in the experimental and simulated samples for quenches above $\Gamma_c$. For this, let us refer to Fig. \ref{snaps3p36}, which considers a sequence of snapshots depicting different stages of the evolution of a cubic monolayer quenched from $\Gamma_0$ towards $\Gamma_f=3.36$ ($f=78$Hz). As illustrated by the three snapshots in the upper panel of this figure, the superheated squared-lattice remains stable over a relatively long time window (roughly, $t=100$ seconds) in spite of the remarkably large value of $\Gamma_f$ considered. One notices that, over this time interval, only a few random dislocations appear (solid circles in Figs. \ref{snaps3p36}b and \ref{snaps3p36}c), arising when some particles escape spontaneously from the monolayer and remain bouncing on top.

Qualitatively, these features may also be contrasted with those found in a hexagonal array of spheres, where the precipitous sublimation of the superheated \emph{crystal} into a granular \emph{gas} is triggered immediately after one single particle 
detaches from the monolayer \cite{pacheco1}. In our present case, instead, we found that the squared-lattice structure remains stable over a remarkably large time window, in spite of the emerging dislocations, and where an increasingly larger number of particles flee the monolayer. The freed space, however, allows to the remaining cubes in the lattice to gradually develop random motion, strongly influenced by rotations \cite{elizondo} and where the particles surrounding the void gain enough momentum from the vibrated plate to distort their local environment. This process, in turn, rapidly results in the germination of multiple clusters of the liquid \emph{phase} (red circles in Fig. \ref{snaps3p36}d), thus signaling the onset of \emph{melting}. Eventually, these clusters grow large enough to become a single and much larger entity, which coexists with the remaining part of the crystal for some time (Fig. \ref{snaps3p36}e). Finally, at a time $t\approx108$s $\equiv\tau_m(\Gamma=3.36)$ the system fully transitions into a homogeneous liquid-like state (Fig. \ref{snaps3p36}f). Remarkably, these features are comparable to some extent to those observed during the solid-liquid transition in thermally driven systems \cite{mura}.

\begin{figure}[h]
\centering
\includegraphics[width=0.242\textwidth, height=0.242\textwidth]{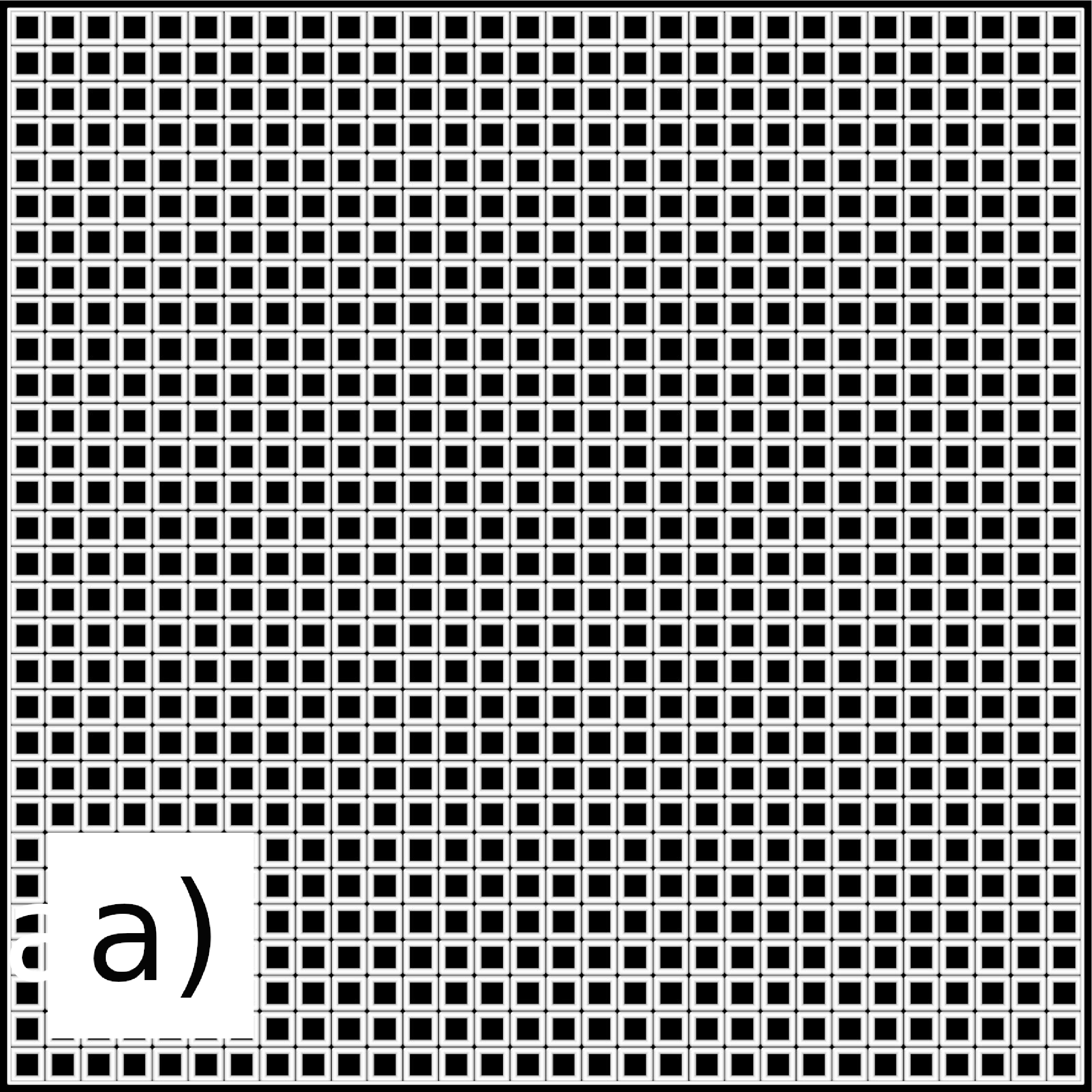}
\includegraphics[width=0.242\textwidth, height=0.242\textwidth]{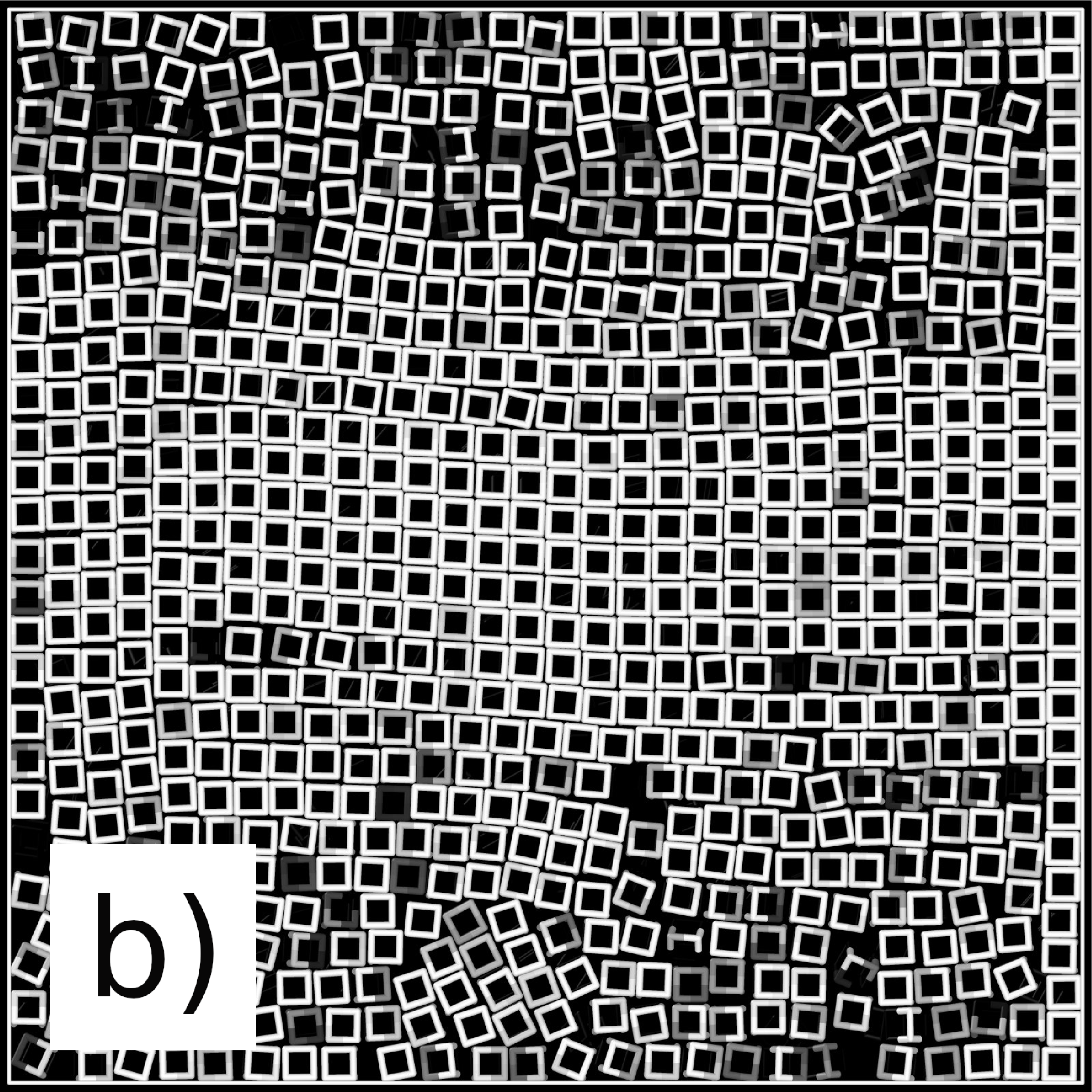}
\includegraphics[width=0.242\textwidth, height=0.242\textwidth]{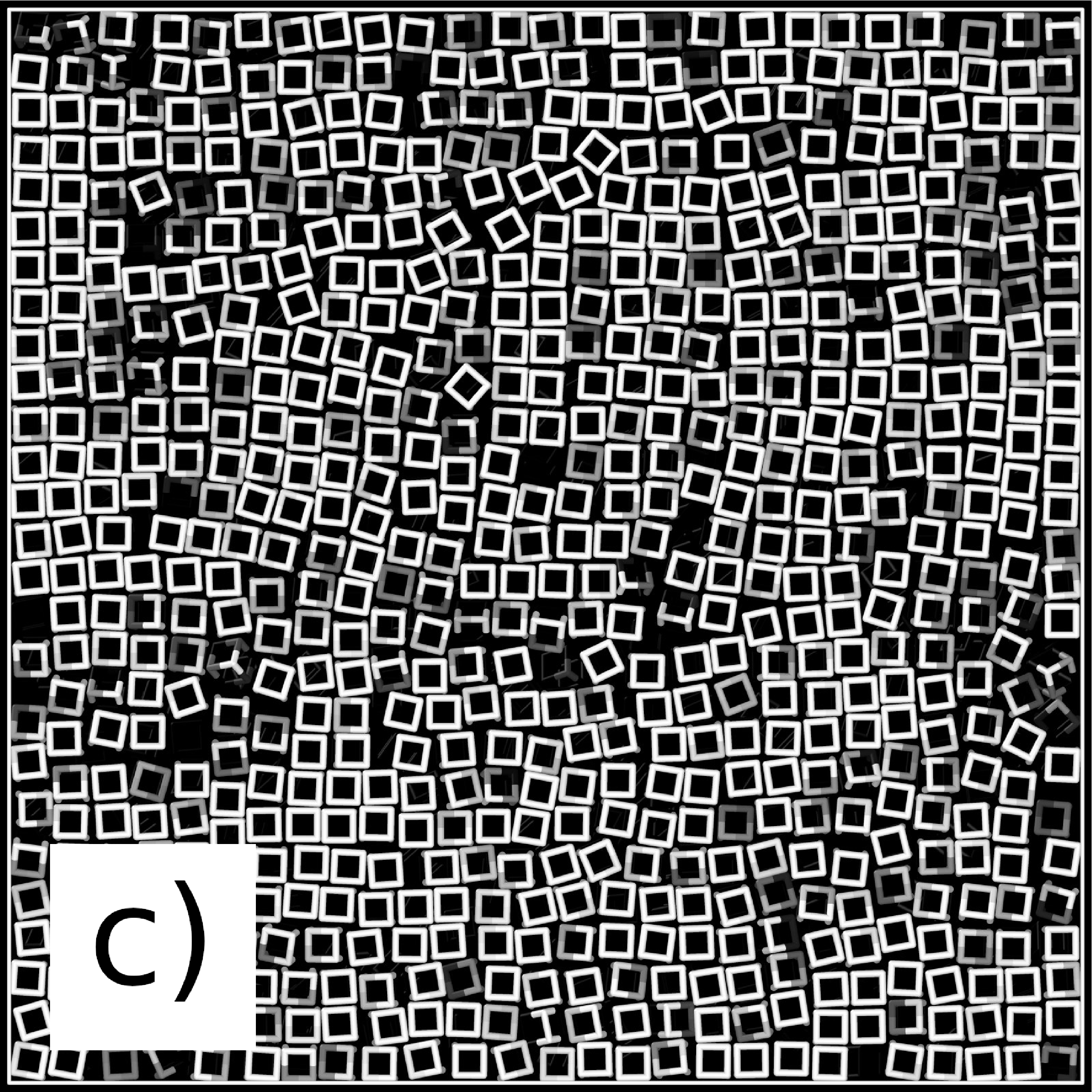}
\includegraphics[width=0.242\textwidth, height=0.242\textwidth]{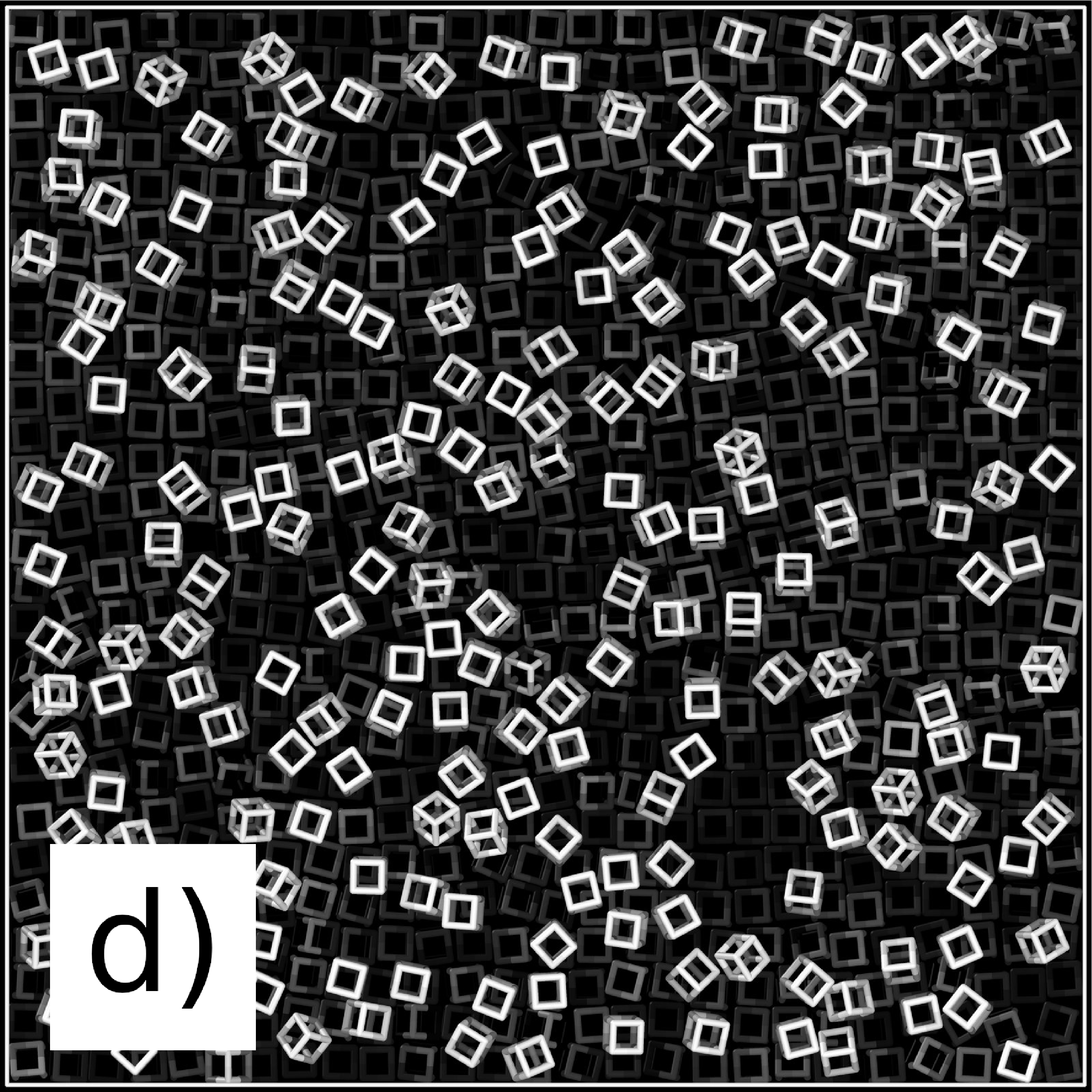}
\caption{Sequence of snapshots describing the evolution of the simulated system of hexahedrons, from the initial state at $t=0$ (a), and describing the gradual melting of the solid, where a transient coexistence of the solid and liquid \emph{phases} can be observed (b and c) before melting (d).} 
 \label{snaps_sim}
\end{figure}

As shown in Figs. \ref{snaps_sim}a-\ref{snaps_sim}c, the simulated system exhibited a melting scenario broadly consistent with the above experimental observations, although with some slight differences that are worth noting. For instance, in our MD simulations we observed that dislocations produced by protruding particles were more likely to appear near the boundaries of the system (although, in some of our runs, they also appeared in the center of the simulation box). Similarly, we noticed that particles adjacent to the confining walls were more prone to remain aligned with them, while a solid in the center of the system melted gradually (Fig. \ref{snaps_sim}b). In all the cases, however, the progressive development of a homogeneous distribution of dislocations (Fig. \ref{snaps_sim}c) preceded the complete transition of the \emph{crystal} state into a \emph{liquid}-like one (Fig. \ref{snaps_sim}d).


\subsection{Radial distribution function}\label{rdf}

\begin{figure}[h!]
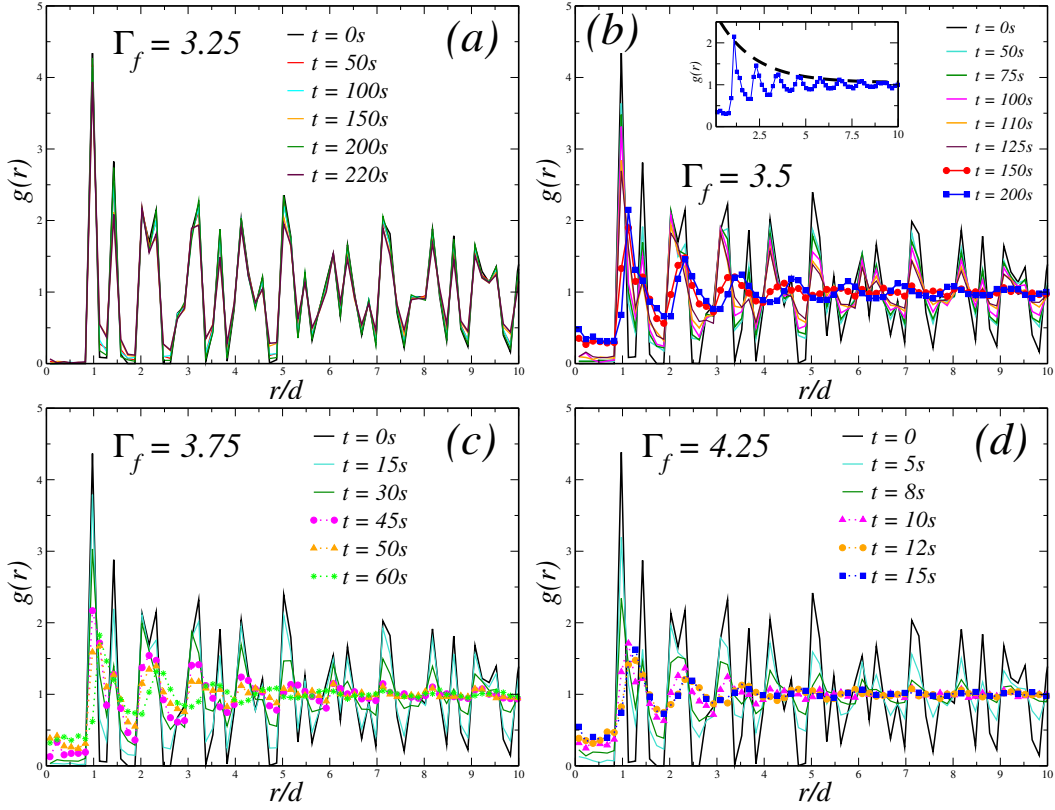

\centering
\includegraphics[scale=0.25]{Gamma3p25.eps}
\includegraphics[scale=0.25]{Gamma3p50_new_new.eps}\\
\includegraphics[scale=0.25]{Gamma3p75.eps}
\includegraphics[scale=0.25]{Gamma4p25.eps}
\caption{Time-evolution of the radial distribution function, $g(r;\tau)$, of the vibrated system of cubic particles after an instantaneous quench towards different values of the shaking strength, namely, (a) $\Gamma_f=3.25$, (b) $\Gamma_f=3.5$, (c) $\Gamma_f=3.75$ and (d) $\Gamma_f=4.25$. Solid lines are used represent crystalline profiles in $g(t;\tau)$, whereas solid symbols account for liquid-like states (see the text). The inset in (b) displays the data for $g(r;t=160s)$ for a quench to $\Gamma_f=3.5$ (solid symbols), along with the function $f(r)=\alpha e^{-\beta r}+C$, which highlights the exponential decay of the RDF maxima (see the text).}  \label{gder}
\end{figure}

Physical insight on the above characteristics of \emph{melting} observed in both, the experimental and simulated systems, can be obtained by analyzing the kinetics of the two-dimensional radial distribution function (RDF) $g(r;t)$, which can be readily determined from the configurations obtained with MD. This canonical observable provides valuable information on the degree of positional ordering in a many body system, and particularly, in the present system \cite{elizondo,frenkel}. Thus, in Fig. \ref{gder} we describe the same transformation process previously illustrated in Figs. \ref{snaps3p36} and \ref{snaps_sim}, but this time, in terms of the $t$-evolution of the RDF for different values of $\Gamma_f$, below (Fig. \ref{gder}\emph{a}) and above (Figs. \ref{gder}\emph{b}-\ref{gder}\emph{d}) the singular value of the shaking strength, $\Gamma^{sim}_c\approx3.255$. The procedure to determine this critical value will be discussed in Subsection \ref{taut}.  

As expected, at the initial state ($t=0$) the RDF displays a regular profile that is characteristic of a 2D tetragonal lattice. More specifically, one notices that the function $g(r;t=0)$ (black solid lines in Figs. \ref{gder}\emph{a}-\ref{gder}\emph{d}) always shows a sequence of sharp and regularly spaced maxima, with very slowly decaying amplitudes over a wide range of $r$, thus indicating long-ranged positional correlations -- or equivalently, a very large correlation length \cite{frenkel}. The first maximum of this sequence, centered at $r=d$, accounts for nearest neighbors located in the direction of the four contact faces of an arbitrary cube. The train of successive peaks, in addition, describes the subsequent shells of neighboring particles in the lattice (e.g. second nearest neighbors, located in the direction of the edges at a center-of-mass distance $r=\sqrt{2}d\approx1.42d$, third nearest neighbors at $r=2d$, etc). 

Hence, in connection with the discussion of the previous Subsection, for quenches with $\Gamma_f<\Gamma_c^{sim}$ one expects that the crystalline profile of the RDF should not display any relevant change at any timescale and, in particular, over the whole time-window accessed with our simulations, $t_{max}=220$s. As illustrated in Fig. \ref{gder}\emph{a}, this is precisely the scenario found for a quench with $\Gamma_f=3.25$, where the superheated crystalline structure remains virtually unaffected over the whole time interval $t_{max}$ considered. 

This is to be contrasted with the scenario obtained upon a relatively small increment of the final shaking strength to $\Gamma_f=3.5$, now above the critical value $\Gamma^{sim}_c\approx3.255$. As shown in Fig. \ref{gder}\emph{b}, in this case a crystalline pattern of the RDF (solid lines) only endures over a time window of, roughly, $t=125$s, along which the height of each maxima -- including the contact value $g(r=d;t)$ -- decreases progressively with $t$. At larger times, the profile of the RDF indicates a homogeneous (\emph{liquid}-like) distribution of particles \cite{elizondo} (solid symbols). One notices that, in this time regime, $g(r<d)\neq0$, due to the contributions of an increasingly larger number of particles that escape from the monolayer and remain bouncing on top. Finally, a full transition into a stationary disordered state is observed at $t=\tau_m(\Gamma_f=3.5)\approx160$s, where the first maximum of the RDF becomes remarkably smaller with respect to the initial state, and shifts towards slightly larger $r$ values. Notice, in addition, that the amplitudes of the subsequent maxima now display a seemingly exponential decay with $r$, thus indicating a finite correlation length. This is emphasized by the inset of Fig. \ref{gder}(b), which displays the stationary value $g(t=200s;r)$ along with the function $f(r)=\alpha e^{-\beta r}+C$ (dashed line), with arbitrarily chosen parameters $\alpha=1.86,\beta=0.5$, and where $C=1$ to represent the boundary condition $g(r\to\infty)=1$.

Qualitatively, the same features are observed in the kinetics of $g(r;t)$ upon increasing $\Gamma_f$ further. One notices, however, that the \emph{melting} timescale $\tau_m(\Gamma_f)$ describing the full crossover from the superheated \emph{crystal} to the \emph{liquid} state, becomes progressively smaller. For $\Gamma_f=3.75$, for example (Fig. \ref{gder}c), we found that $\tau_m(\Gamma_f=3.75)\approx60$s (i.e. almost a third of the value found in the previous case) and, for $\Gamma_f=4.25$, $\tau_m(\Gamma_f=4.25)\approx15$s (Fig. \ref{gder}(d)). One observes, in addition, that the final stationary profile of the RDF also depends on the value of $\Gamma_f$, showing increasingly smaller peaks, and shorter correlation lengths, for larger $\Gamma_f$. Let us finally mention that, overall, the same trends were consistently found in the experiments and simulations at the level of $\tau_m(\Gamma)$, with only small quantitative differences. 

\subsection{Dependence of the melting time $\tau_m$ on $\Gamma_f$ and connection with previous work}\label{taut}

A key insight from the MD results of Fig. \ref{gder}, fully consistent with our main experimental findings, is the existence of a critical value $\Gamma_c$ separating the two distinct regimes for the evolution of the cubic monolayer. In the domain $1<\Gamma_f<\Gamma_c$, superheated states that prevail indefinitely are obtained and, thus, they are described by the condition $\tau_m\to\infty$. In the complementary domain $\Gamma_f>\Gamma_c$, instead, $\tau_m$ is finite and becomes increasingly smaller with larger $\Gamma_f$. Hence, a relevant issue concerns to the determination of $\Gamma_c$, and, as we will show in what follows, this can be achieved by analyzing the dependence of $\tau_m$ on $\Gamma_f$.

With this goal in mind, let us highlight first that the above physical scenario bears qualitative similarities with that predicted theoretically for the solid-liquid phase transition in a two dimensional system of planar hard-squares, a rigid polyhedral shape with a four-fold symmetry \cite{nelson}. Such transition, in fact, has also been investigated with the assistance of MD simulations \cite{frenkel,avendano}, and also with experiments on dense colloidal suspensions of hard-square \cite{mason,keim} and kite-shaped platelets \cite{hou}. In some of these studies, the melting temperature $T_m$ separating the solid phase from the liquid is determined empirically from the assumption that the correlation length $\xi$ -- defined through a relation of the form $g(r)=\exp{(r/\xi)}$ -- diverges exponentially as one approaches $T_m$ from above. More specifically, such divergence is assumed to be described by the function $\xi(T)\sim \exp\left\lbrace B(T-T_m)^{-\nu}\right\rbrace$, where $B$ is an arbitrary constant, and with an exponent $\nu\approx 0.36963$ for hard-squares (see, for instance, Eq. (10) of Ref \cite{frenkel}, or Eq. (1.10a) of Ref. \cite{nelson}). One might be tempted, thus, to adopt a similar line of reasoning, and assume an analogous dependence of $\xi$ on $\Gamma_f$, from which one could approximate the critical shaking strength $\Gamma_c$ from a fit of experimental and simulated data for the RDF. 

There are various reasons for which we considered this route not suitable. The first of them, concerns to the practical difficulties to resolve experimentally particle trajectories in the $(x,y)$-plane for a vibrated system of cubes, in the absence of vertical confinement and at large concentrations \cite{elizondo}. This renders the experimental determination of the RDF particularly complicated, which thus hinders the comparison with simulated results. More importantly, however, the previous results indicate that the superheating behavior of the present granular system has a clear kinetic nature and, thus, can hardly be understood by only appealing to phase-transition arguments. Furthermore, in view of the scarcity of studies on granular superheating phenomena, it would be highly desirable to compare our findings, at least qualitatively, with the few available results, particularly, those reported in Ref. \cite{pacheco1}. 

\begin{figure}[h]
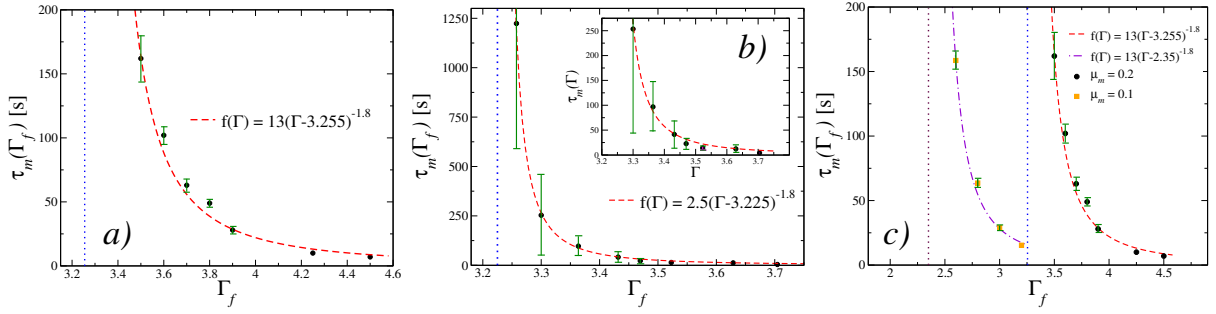

\centering
\includegraphics[width=.315\textwidth]{fig5a_final.eps}
\includegraphics[width=.315\textwidth]{fig5b_final.eps}
\includegraphics[width=.315\textwidth]{LASTPLOT.eps}
\caption{Average melting time, plotted as a function of the shaking strength $\Gamma_f$ (solid circles) and obtained via MD simulations (a) and experiments (b). In both cases, the (red) dashed lines describe a power-law relation of the form $\tau(\Gamma)=A(\Gamma-\Gamma_c)^{-\nu}$, used to fit the simulated and experimental data for $\tau_m(\Gamma)$, and yielding the fitting parameters $A^{sim}=13$, $\Gamma_c^{sim}=3.255$ and $\nu=1.8$ for the MD simulations, and $A=2.5$, $\Gamma_c=3.225$ and $\nu=1.8$, for the experiments (see the text). The inset in (b) is a close-up of a time-window comparable to that explored with MD. (c) Average melting times found in the simulated system by decreasing by half the friction coefficient $\mu_m$ (solid squares). The data of (a) is re-plotted for reference (circles). The two dashed lines, in turn, describe power laws that only differ in the value $\Gamma_f$ (see the text).}
\label{tau_gamma}
\end{figure}

Therefore, rather than considering the correlation length $\xi(\Gamma_f)$, we shall employ the \emph{melting} time $\tau_m(\Gamma_f)$ as the physical parameter to estimate the location of the critical value $\Gamma_c$, since this quantity can be easily determined from both, the recorded evolution of the experimental system and the MD simulations. Specifically, here we stick to the proposal of \cite{pacheco1}, which assumes a power-law dependence $\tau_m(\Gamma_f)=A(\Gamma_f-\Gamma_c)^{-\nu}$, allowing to fit our results for the melting time as a function of $\Gamma_f$. Such data are displayed in Figs. \ref{tau_gamma}a and Fig. \ref{tau_gamma}b (solid symbols) for, respectively, simulations and experiments. For reference, in the same figures we also plot the aforementioned power laws (dashed lines).

Referring first to the results in Fig. \ref{tau_gamma}a, one notices that the $\Gamma$-dependence of the (average) \emph{melting} time $\tau_m$ (solid circles) obtained with our simulations is well described by the fit proposed, using the fitting parameters $A^{sim}=13$, $\Gamma_c^{sim}=3.255$ and $\nu=1.8$. Despite being constrained by the maximum accessible simulation time ($t_{max}=220$s), these results
provide a reasonable estimate for the critical shaking strength $\Gamma_c$, and  more crucially, they also reveal a remarkable feature regarding the superheated metastable states of the present cubic system. We refer to the fact that the numerical value of the exponent $\nu$ used in our fit, is virtually the same as that obtained for a hexagonal monolayer of spherical particles, namely, $\nu^{sph}\approx1.7$ (see Fig. 2b of Ref. \cite{pacheco1}). This coincidence, thus, might suggest the universality of the exponents describing the lifetime of superheated states in driven granular matter.

Such conjecture seems to be supported by the experimental data obtained for the plot of $\tau_m$ vs $\Gamma_f$, presented in Fig. \ref{tau_gamma}b (symbols), and which are well described by the same power-law used to fit our simulation data. The (red) dashed line displayed in this figure, for example, was obtained using the fitting parameters $A=2.5$, $\Gamma_c=3.225$ and $\nu=1.8$.
Remarkably, except for the difference in the numerical value of the coefficients $A^{sim}$ and $A$, we obtain comparable values for the critical shaking strength and, more importantly, the same superheating exponent as before. We attribute the numerical deviations to the intrinsic differences between the simulated and experimental systems and, possibly, to the two distinct paths to vary the parameter $\Gamma_f$. 

To further examine the possible universality of the superheating exponent $\nu$, we conducted additional simulations in which the friction coefficient $\mu_m$, characterizing the dissipative contributions to the interactions between the simulated particles, was reduced by half. We then determined the corresponding dependence of the melting time $\tau_m$ on the shaking strength $\Gamma_f$. The results, presented in Fig. \ref{tau_gamma}(c) (solid squares), reveal that the simulated data (averaged over five realizations) adheres to the same power-law as before, with the only difference being a downward shift in the critical value $\Gamma_c$. This finding suggests that, while the location of the transition is sensitive to frictional dissipation (as expected), the scaling behavior of the melting time $\tau_m$ remains invariant. Note that this also indicates that the parameter A in our fit is independent of friction. 


Unfortunately, we cannot say much at the moment on the nature of the exponents $\nu$ and $\nu^{sph}$, whose first principles description actually remains as a fundamental challenge for modern statistical physics. However, the fact that essentially the same scaling laws describe the lifetime of superheated states observed in two granular systems comprised by very dissimilar particles (cubic and spherical), allows for some additional speculations. Let us elaborate.

As discussed in Ref. \cite{pacheco1}, one intuitively expects that the critical behavior represented by the above superheating exponents must be linked to the energy dissipation mechanism, in such way that the time scale $\tau_m$ is proportional to the size of dissipating regions in the crystal. At large $\Gamma_f$ $(>\Gamma_c)$, for example, these regions are expected to decrease very fast, since the energy supplied by the vibrating plate detaches easily many particles from the monolayer, producing multiple dislocations in the squared-lattice. This allows to the particles surrounding a void to dissipate less energy in the form of face-to-face and face-to-bottom contacts; and instead, to gain more kinetic energy from the oscillating plate, thus dismantling the ordered distribution of cubes easily. This shortens the melting time $\tau_m$.


Conversely, by approaching the critical value $\Gamma_c$ from above, the propensity of the cubes to escape the monolayer is much lower, and we observed that a few dislocations do not affect crucially the squared-lattice distribution of particles over a long time. Hence, the predominant configurations for contact interactions remain to be face-to-face (among the cubes) and face-to-bottom (cubes and the bottom plane) collisions, which are optimal to dissipate the energy supplied. In other words, as $\Gamma_f$ approaches $\Gamma_c$ from above, the system can dissipate such energy more efficiently, which thus increases the time scale $\tau_m$.  

\begin{figure}[h]
\centering
\includegraphics[scale=0.45]{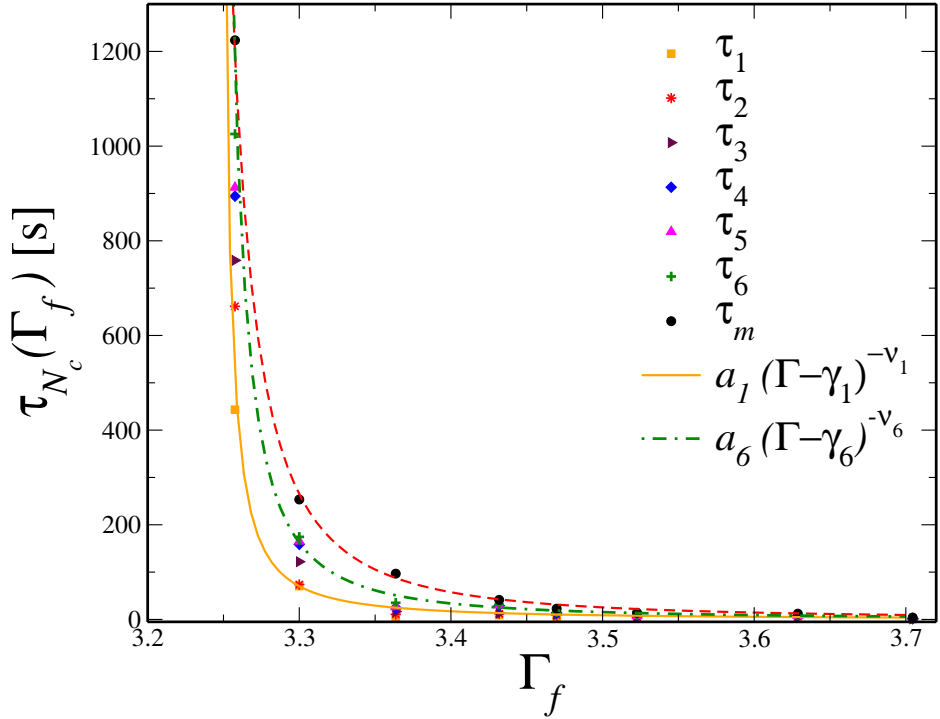}
\caption{(a) Mean time $\tau_{N_c}$ (solid symbols) required by a number $N_c$ of cubes (as indicated, $N_c=1,2...6$) to escape the monolayer, plotted as a function of the shaking strength $\Gamma_f$. For reference, we also display the data of Fig. \ref{tau_gamma}(b) for the \emph{melting} time $\tau_m(\Gamma)$ and the power law $f(\Gamma)=2.5(\Gamma-3.225)^{-1.8}$. The solid and dashed lines are fits for the experimental data for $\tau_{N_c=1}(\Gamma)$ and $\tau_{N_c=6}(\Gamma)$, using the fitting parameters $a_1=1.4$, $\gamma_1=3.245$, $\nu_1=1.4$, and $a_6=1.8$, $\gamma_6=3.24$, $\gamma_6=1.6$, respectively (see the text).}
\label{dice_num}
\end{figure}

Whit this simplified picture in mind, it is reasonable to surmise that the number of cubes escaping from the monolayer provides an indirect measure of the lost of dissipation regions during the melting process of the superheated crystal, allowing to test the validity of the above arguments (at least qualitatively) by monitoring the number of protruding cubes as a function of time, for different shaking strengths. Fig. \ref{dice_num} summarizes our experimental results for the (average) time, $\tau_{N_c}$, required by a (variable) number $N_c$ of cubes to flee the monolayer, plotted as a function of $\Gamma_f$ (solid symbols). As just explained, for sufficiently large $\Gamma_f$ $(>3.45)$ all the cubic particles in the monolayer gain enough energy from the vibrating plate to escape with ease, and hence, the melting of the squared-lattice occurs within a remarkably short time-window. For $\Gamma_c<\Gamma_f<3.45$, instead, the occurrence of each output-event becomes increasingly spaced out in time by approaching the critical value $\Gamma_c$ from above, so that one can monitor easily each of the output-events.

Surprisingly, for all the relevant values of $\Gamma_f$ considered here, we observed approximately the same geometrical condition for the onset of melting, namely, that six cubes have escaped the monolayer, or $N_c=6$. We noticed that, when this condition is fulfilled, isolated nucleation domains of the \emph{liquid}-phase are formed and, later, these domains percolate to form a larger entity. Let us highlight that this condition corresponds to a relatively small lost in the dissipation regions of the lattice, which can be accounted for using the quantity $\delta\equiv 5d^2\cdot(N_c)/5d^2\cdot(900)$, that considers the ratio of the (ideal) dissipation region lost after $N_c$ cubes detach from the monolayer, to the (ideal) dissipation region of the initial configuration. Thus, for $N_c=6$, $\delta\approx0.007$. It may be of some value to highlight that this number is at least comparable to that corresponding to the hexagonal lattice of vibrated spheres reported in \cite{pacheco1}, in which the condition observed for an almost instantaneous sublimation is that a single sphere (with, ideally, seven point contacts) escapes from the monolayer, thus corresponding to a ratio of dissipation regions of $\delta\approx0.009$. 

Let us also highlight that each of the plots of $\tau_{N_c}$ vs $\Gamma_f$ ($N_c=1,...6$) obtained, can be well fitted used the same power law dependence as before, namely, $\tau_{N_c} = a_{N_c} (\Gamma_f-\gamma_{N_c})^{-\nu_{N_c}}$ (solid and dashed-dotted lines in Fig. \ref{dice_num}), with an exponent $\nu_{N_c}$ that depends on $N_c$. As the number of protruding cubes increases, one observes that the plot of $\tau_{N_c}$ vs $\Gamma_f$ approaches gradually to that of $\tau_m(\Gamma_f)$ and, consequently, the exponent $\nu_i$ approaches the value $1.8$ of the superheating exponent $\nu$. For example, for $N_c=1$, we found that the parameters $a_1=1.4$, $\gamma_1=3.245$ and $\nu_1=1.35$ (solid line) yield a nice agreement with our results for $\tau_1(\Gamma_f)$ (solid squares). Similarly, for $N_c=6$, the fitting parameters $a_6=1.8$, $\gamma_6=3.24$ and $\nu_6=1.6$ (dashed-dotted line) provide a consistent description of our data for $\tau_{6}(\Gamma_f)$

\subsection{Melting profiles and the Johnson-Mehl-Avrami-Kolmogorov model}\label{prof}

We finally comment on another interesting feature observed in the transition from the superheated \emph{solid} to the \emph{liquid} state in our granular cubic system. For this, let us recall here a well established approach for the description of the progress of phase transformations in material systems, namely, the so-called Johnson–Mehl–Avrami–Kolmogorov (JMAK) theory \cite{avrami,shirzad}. This approach, originally developed in the context of the classical nucleation theory, is summarized by a sigmoidal equation of the form $\phi(t)=1-\exp(-kt^\gamma)$, in which $\phi(t)$ is the fraction of material that , at time $t$, has been transformed into the new phase, and $k$ and $\gamma$ are system's specific constants. In spite of its apparent simplicity, the JMAK equation has been employed successfully in a variety of contexts, including the description of the kinetics of non-thermodynamic transformations \cite{shirzad}.

\begin{figure}[h]
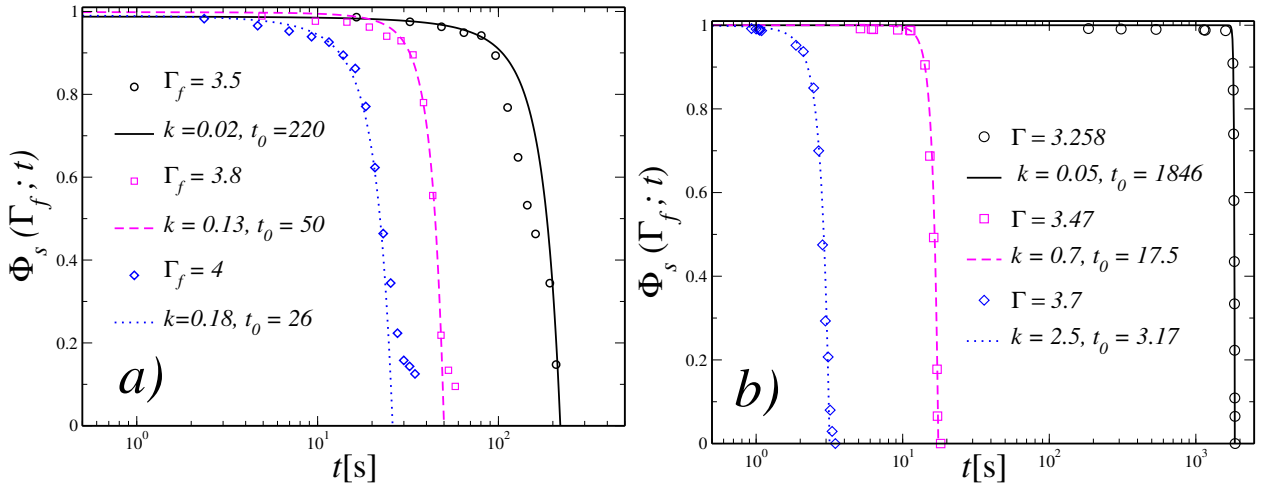

\centering
\includegraphics[width=.495\textwidth]{fig8a.eps}
\includegraphics[width=.495\textwidth]{planar_density_new.eps}
\caption{Evolution of the solid fraction $\Phi_s(t)$ (open symbols) as obtained from simulations (a) and experiments (b). The dotted, dashed and solid lines in both panels describe the JMAK-like equation of the form $h(t)=1-\exp(k(t-t_0))$, using $k$ and $t_0$ as fitting parameters, as indicated.}  \label{metastabletimes}
\end{figure}

Hence, as a mere exploratory exercise we verified whether the processes of transformation observed in our system could be accounted for by this kind of equation, with the aim of exhibit its applicability to the description of \emph{phase} transformations at the granular level. For this, we analyzed various \emph{melting} processes for different $\Gamma_f$, along which we monitored the $t$-evolution of the remaining fraction of the \emph{solid} phase, $\Phi_s(t)$ (rather than the fraction $\phi(t)$ of the \emph{liquid}, for practical convenience). Our results obtained with simulations \cite{simmi} and experiments \cite{expi} are displayed, respectively, in Figs. \ref{metastabletimes}a and \ref{metastabletimes}b (open symbols), for three different values of the shaking strength, as indicated. 

For all the \emph{melting} processes considered, we observed a slow speed of transformation at the beginning, where $\Phi_s(t)$ remains nearly constant. This is consistent with the observation that a number of cubes $N_c=6$ must detach from the monolayer first, allowing for the formation of voids that,  eventually, incubate the nucleation domains. As already discussed, far above the critical value $\Gamma_c$ this condition is achieved relatively faster, so that the solid fraction $\Phi_s$ relaxes exponentially with $t$ after a time-window of a few seconds. By decreasing $\Gamma_f$, however, the capacity of a cube leaving the monolayer is lower and, thus, $\Phi_s$ remains constant over larger times, but showing eventually the same exponential relaxation profile. As one comes closer to the critical value $\Gamma_c$, in addition, the time required for the full decay of the solid-fraction increases dramatically (two decades in our simulations, almost three decades in our experiments).


As illustrated by the dotted, dashed and solid lines in Fig. \ref{metastabletimes}, a JMAK-like equation of the form $h(t)=1-\exp(k(t-t_0))$ (i.e. with exponent $\gamma=1$) describes almost perfectly the decay of $\Phi_s(t)$ obtained with our simulations \cite{simmi} and experiments \cite{expi}, using $k$ and $t_0$ as fitting parameters (as indicated). Interestingly, within the JMAK description of material transformations this kind of behavior is interpreted as follows: An initial slow transition step starts with the spontaneous emergence of small nucleation domains. This stage is followed by a period of rapid growth of the new phase, due to the increasing availability of the inter-facial area between the solid and the liquid, allowing to an increasingly larger number of particles to detach from the "parent" phase and conform the "product" one. The process ends when the system runs out of transformable portions of material. In essence, this is the scenario outlined so far by our simulations and experiments, hereby confirmed also at the level of $\Phi_s(t)$.




A few final remarks might be in order here. Within the JMAK model, the ``\emph{constant}'' $k$ is in reality a function of both, nucleation and growth rates, whereas the exponent $\gamma$ is typically related to the dimensionality of growth \cite{avrami,shirzad}. Since, in our fit $k$ becomes smaller approaching $\Gamma_c$, both the nucleation and growth rates must become increasingly smaller close to the critical point. This is consistent with our main experimental observations. Furthermore, the fact that the exponent $\gamma$ equals unity in our fit, suggests that the \emph{melting} scenario observed in the present granular system resembles closely the transformation process referred to as pĺate-like diffusion-controlled growth \cite{avrami,shirzad}. These aspects will be analyzed in more detail in future work.

\section{Concluding Remarks}\label{conclu}

In this work, we combined experiments and molecular dynamics simulations to develop a consistent description of superheating phenomena in driven granular matter. As a model system, we employed cubic particles initially arranged in a square-lattice monolayer, and subjected to vertical vibrations at various shaking strengths $\Gamma_f$, both below and above a critical threshold $\Gamma_c$. In qualitative agreement with previous observations on hexagonal monolayers of spherical grains \cite{pacheco1}, we found that for $\Gamma_f<\Gamma_c$, superheated-\emph{solid} states persist indefinitely, while for $\Gamma_f>\Gamma_c$, the system undergoes a transition to a fluid (\emph{liquid}-like) state. This transition occurs over a characteristic timescale $\tau_m(\Gamma_f)$ that diverges as one approaches $\Gamma_c$ from above. 

Clearly, the geometry of the cubic particles leads to a significant increase in frictional contact area, relative to that of a monolayer of spherical particles, thus enhancing the ability of a granular system to dissipate the energy supplied via vibrations. This leads to a marked upward shift of the critical shaking strength, as compared to that found in Ref. \cite{pacheco1}, and also, contributes to extended the stability of the transient superheated states. These findings, thus, highlight the role of friction and geometry in controlling the durability of superheated granular assemblies, which parallels the effect of the surface coatings used in crystalline solids to inhibit nucleation and delay melting, (i.e. to extend the superheated regime). 

The analysis of the kinetics of the (two dimensional) radial distribution function reveals that the present system of cubic particles exhibits some similarities to the solid-liquid transition observed in systems of hard squares, in which the (spatial) correlation length evolves, from a power-law decay in the solid phase, to an exponential decay in the liquid. In our present case, however, the absence of two dimensional confinement precludes the comparison with such scenario, since the nature of the superheated-\emph{solid}-to-\emph{liquid} transition is governed by different underlying mechanisms. Specifically, \emph{melting} is initiated by the escape of individual cubes from the monolayer, introducing dislocations in the ordered structure. The space freed up by the protruding particles, allows to the remaining cubes to develop gradually out-of-plane rotations and, hence, to gain sufficient energy from the vibrating plate to develop random (Brownian-like) motion \cite{elizondo}. In view of the kinetic pathway for this process, the melting time $\tau_m$ serves as a more appropriate and physically meaningful parameter to characterize the transition from the superheated \emph{solid}-like state to the disordered \emph{liquid}-like regime.

By determining systematically this quantity in simulations and experiments, we uncovered striking features. First, and despite intrinsic differences in particle composition and interaction details, both approaches yield consistent estimates for the critical shaking strength $\Gamma_c$. More importantly, however, we find that the divergence of $\tau_m$ in the approach to the critical value follows the same power-law in both cases, with virtually the same exponent $\nu=1.8$. Remarkably, this exponent appears to be largely insensitive to frictional effects, as exhibited here by additional simulations in which the friction coefficient $\mu_m$ was reduced by half, without altering the scaling behavior. These features, hence, suggest the possible universality of the superheating exponent $\nu$, a possibility further supported by the findings of Ref. \cite{pacheco1}, where essentially the same power-law divergence was observed for the sublimation of a hexagonal monolayer of spherical particles.

In this regard, another remarkable finding concerns the geometrical condition observed experimentally at the onset of \emph{melting}. Specifically, we found that the transition is triggered when the number of protruding cubes reaches the value $N_c=6$, irrespective of $\Gamma_f$. This suggests that a loss of a dissipation area equivalent to the five contact faces per detached cube -- amounting to a total of 30 missing contact faces -- suffices to trigger the nucleation of the liquid phase. Using oversimplified arguments for an ideal jammed state, in which all the cubes maintain full face-to-face and face-to-bottom contacts, one can easily estimate the dissipation area lost as a function of $N_c$. Interestingly, the numerical result for $N_c=6$ is comparable to that obtained for the ideal dissipation area lost in a hexagonal monolayer of spheres, where the condition that marks the onset of the superheated-\emph{solid}-to-\emph{gas} transition is $N_{sph}=1$ \cite{pacheco1}.


Moreover, when the melting processes observed in our simulations and experiments are analyzed through the temporal evolution of the remaining solid fraction, $\Phi(t)$, we find that the data are remarkably well described by a simple expression inspired by the Johnson-Mehl-Avrami-Kolmogorov (JMAK) model -- an approach widely used to characterize phase transformations across various equilibrium and out-of-equilibrium systems. While further investigation is needed, this finding suggests that key concepts from JMAK theory may be transferable to the study of non-equilibrium transitions in driven granular matter. Exploring this potential correspondence opens an intriguing avenue for future research, which we intend to pursue.

The present work complements and extends previous work \cite{elizondo} focused on the \emph{gas} and \emph{liquid} phases of the present granular system. Hence, our present contribution provides additional support to the interpretation that a system comprised by cubes, and subjected to mechanical stimuli, can be employed as an amenable prototypical model to investigate a plethora of (equilibrium and non-equilibrium) in thermally driven systems.  




\end{document}